\documentclass[sigconf,nonacm]{acmart}

\renewcommand\footnotetextcopyrightpermission[1]{}
\usepackage{graphicx}
\usepackage{booktabs}

\begin{document}

\title{What Could the Agent See at 19:05? \\ Generating Temporal Enterprise Scenarios from Real Research and Replaying Them to Evaluate Agents}

\author{Tezan Sahu}
\affiliation{%
  \institution{Microsoft}
  \city{Hyderabad}
  \country{India}
}
\email{tezansahu@microsoft.com}

\author{Himani Arora}
\affiliation{%
  \institution{Microsoft}
  \city{Hyderabad}
  \country{India}
}
\email{hiarora@microsoft.com}

\begin{abstract}
Enterprise AI agents act across many apps whose data changes continuously, so an answer is correct only relative to \emph{what data existed} and \emph{who could see it} at the moment it was asked. Offline evaluation today grades against a single \textbf{static snapshot}---effectively the end of the episode---so it can only evaluate one situation, the \emph{final} one, even though every earlier moment of the episode is a different situation that invites its own realistic questions with its own correct answers. Recreating each of those moments as a separate snapshot would mean re-provisioning a whole tenant per instant, which is prohibitively costly; and even a single snapshot leaks future state hidden \emph{inside} records and cannot represent the multi-app, time-ordered way real work happens. \textbf{Our system} closes two gaps at once: it \emph{generates} a realistic, persona-driven, temporally-evolving enterprise world from real research, and \emph{replays} that world at any chosen moment to evaluate any pluggable agent. A schema-inferred temporal description drives a deterministic-plus-LLM rebuild of each record's past state; because the queryable moments are finite, all rebuilds are precomputed into a compact difference cache, making evaluation a fast, reproducible lookup with no model in the path. We describe the design, an architecture spanning both flows, and early experience evaluating enterprise agents.
\end{abstract}

\maketitle

\begin{figure*}[t]
\centering
\includegraphics[width=0.84\textwidth]{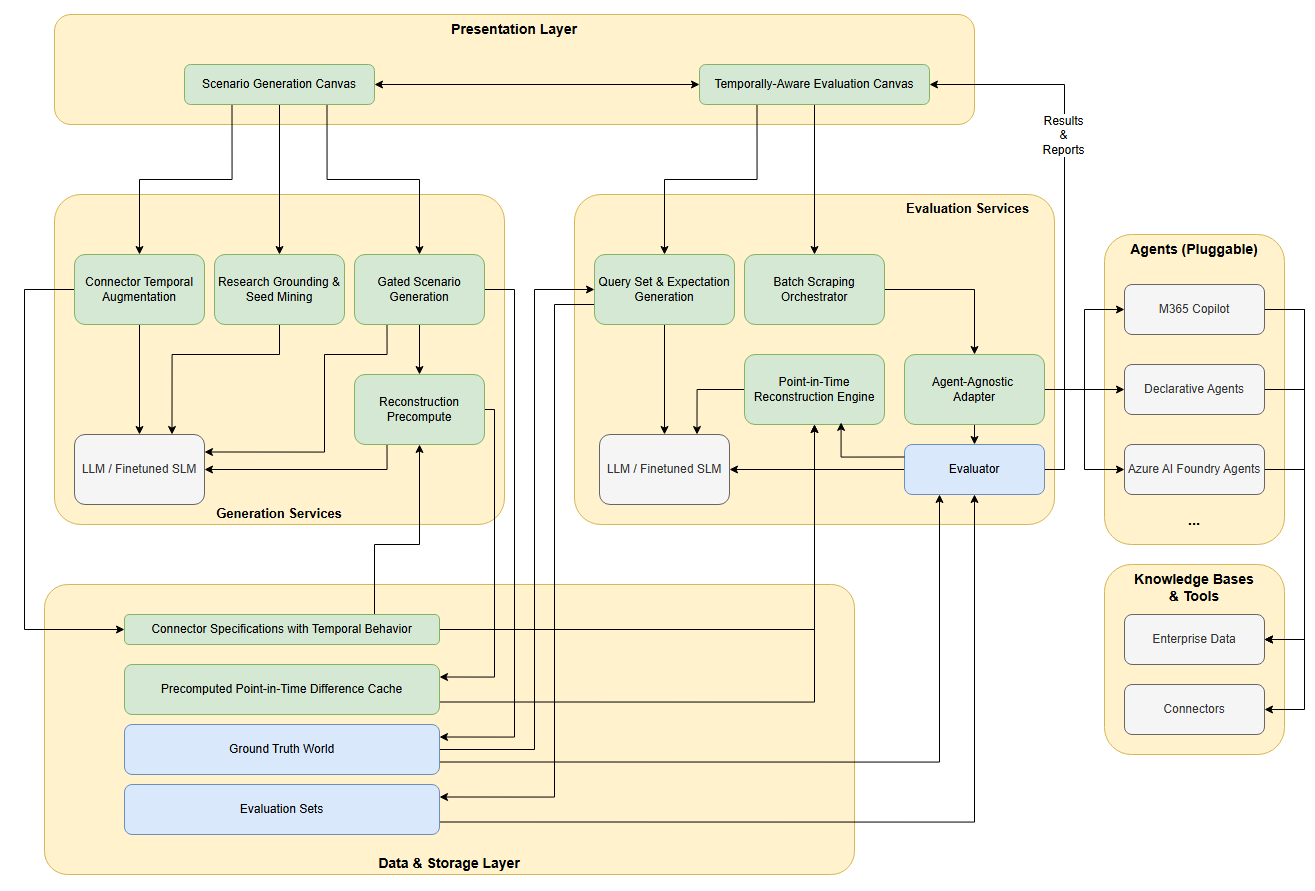}
\caption{System architecture, spanning both flows. \textbf{Generation} mines real research into grounded scenario seeds and expands them, under user-approval gates, into personas, a narrative arc, a timestamped timeline, and cross-referenced artifacts across every app---the full, all-time ground truth---while \emph{temporal augmentation} infers each app's temporal behavior from its schema and a \emph{precompute} step stores point-in-time differences. \textbf{Evaluation} reconstructs the access-scoped world at a chosen moment by cache lookup and merge (no model in the path), runs any agent through a uniform adapter, and grades it against machine-generated, moment-anchored expectations.}
\label{fig:arch}
\end{figure*}

\section{When the Evaluator Cannot Rewind}

Software is beginning to do open-ended knowledge work on people's behalf. A new class of AI \emph{agents}---like M365 Copilot~\cite{copilotextend}, together with the cross-app \emph{skills}, and role-specific \emph{plugins} now shipped by every major vendor~\cite{anthropicfinance,openaiworkspace,agentskills}---reads a person's email, chat, files, and connected business systems (an issue tracker, a code repository, an incident tool) and acts across all of them to carry out a task. Such an agent is only as valuable as it is trustworthy, and the competitive question for these products is not whether they can be built but whether that trust can be established \emph{before} release. Trust is earned through evaluation: the agent is replayed over realistic situations and its answers are graded against what was actually correct. Evaluation is the gate---the more rigorous it is, the better the agent that survives it.

This is, in software-engineering terms, acceptance and regression evaluation, but of an unusually hard kind. The system under evaluation is nondeterministic (a language model), and its ``input'' is not a request payload but an entire \emph{world of evolving, access-controlled enterprise data} that the agent reads and writes through many apps. Producing faithful evaluation data for such a system is the central difficulty---and the way the industry does it today does not hold up.

\textbf{The static-snapshot trap.} Standard practice stands up a \emph{synthetic tenant} (a sandbox of fabricated users and data), freezes it into one \emph{snapshot}, and grades the agent against that single state~\cite{foundryevals}. But real work is not a state; it is a \emph{story} that unfolds over hours. Each moment of that story is a \emph{different situation}---mid-triage, fix-under-review, just-resolved---and each situation invites its own realistic questions with its own correct answers, asked by whichever person is involved. A single snapshot captures exactly one situation, the last, so every other situation the episode passed through is simply not evaluable, and an agent is graded as though every story were already over. One could imagine standing up a fresh snapshot for each moment worth evaluating, but that means re-provisioning a tenant, its connectors, and its mock data for every instant---dozens of near-identical tenants per episode---which is prohibitively expensive.

Consider a concrete episode (the kind Fig.~\ref{fig:gen} builds). A payments service degrades; over two hours an incident is opened, an on-call engineer acknowledges it, a configuration change is suspected, a fix is opened as a pull request and merged, and the incident is finally resolved with a written root cause. At 19:05---twenty-three minutes in, mid-triage---the situation is its own: the engineer naturally asks the agent, \emph{``what is the status of the incident I'm leading, and what's on the timeline so far?''} The true answer then is \emph{acknowledged}, three timeline entries, no root cause yet. But the tenant snapshot is the \emph{end} of the story, so the agent is graded against a world where the incident is already \emph{resolved}, with a full timeline and root cause (Table~\ref{tab:example}). The agent that correctly says ``acknowledged'' is marked \emph{wrong}; the one that parrots the resolved end-state is marked \emph{right}. Evaluation rewards exactly the wrong behavior---and the entire span of mid-episode situations, each with different right answers, never gets evaluated at all.

\begin{table}[h]
\small
\centering
\caption{The same incident record at 19:05 (the truth) versus in the end-state snapshot the agent is actually graded against. The snapshot rewards the wrong answer.}
\label{tab:example}
\begin{tabular}{@{}lll@{}}
\toprule
\textbf{Field} & \textbf{Truth at 19:05} & \textbf{End-state snapshot} \\
\midrule
status & acknowledged & resolved \\
timeline & 3 entries & 10 entries \\
root cause & (none yet) & full text \\
resolved-at & --- & 20:55 \\
\bottomrule
\end{tabular}
\end{table}

Four failures follow from this one root cause, and together they make snapshot evaluation untrustworthy. \emph{(i) Only the final situation can be evaluated.} A snapshot is one state, so the many distinct situations an episode passes through---each with its own natural questions and answers---collapse to just the last one, and the rich \emph{middle}, where most real questions are asked, is unreachable. \emph{(ii) The future leaks from inside records.} A natural patch is to drop records ``newer than $t$,'' but state lives \emph{inside} records too: a ticket accrues later comments, an incident accrues a resolution, a pull request eventually merges. Keep the record but fail to trim its interior, and tomorrow's facts leak into today. \emph{(iii) Hand-built ground truth does not scale.} Authoring even one believable, internally consistent, multi-app episode---and its expected answers---takes a skilled person \emph{days}, and must be redone for every person, moment, and app, so coverage stays tiny. \emph{(iv) An entire class of agents cannot be evaluated.} Increasingly, agents are \emph{always-on}: their value is to \emph{react to a change}---``flag me when a dependency I rely on slips.'' Evaluating that requires placing the agent at the instant \emph{just after} the change; a snapshot, showing only the final state, makes it impossible in principle. (Resetting a system to a clean baseline is well understood; what is missing is stepping \emph{back into the middle} of a multi-app world.)

The remedy needs two things no current harness provides together: a way to \emph{manufacture} a realistic, time-evolving, multi-app world cheaply (Flow~1), and a way to \emph{step into} it at any instant, as any person, to evaluate an agent there (Flow~2). The system is built around exactly these two flows (Fig.~\ref{fig:arch}).

\section{Flow 1: Manufacturing a Believable, Evolving World}

\begin{figure*}[t]
\centering
\includegraphics[width=0.8\textwidth]{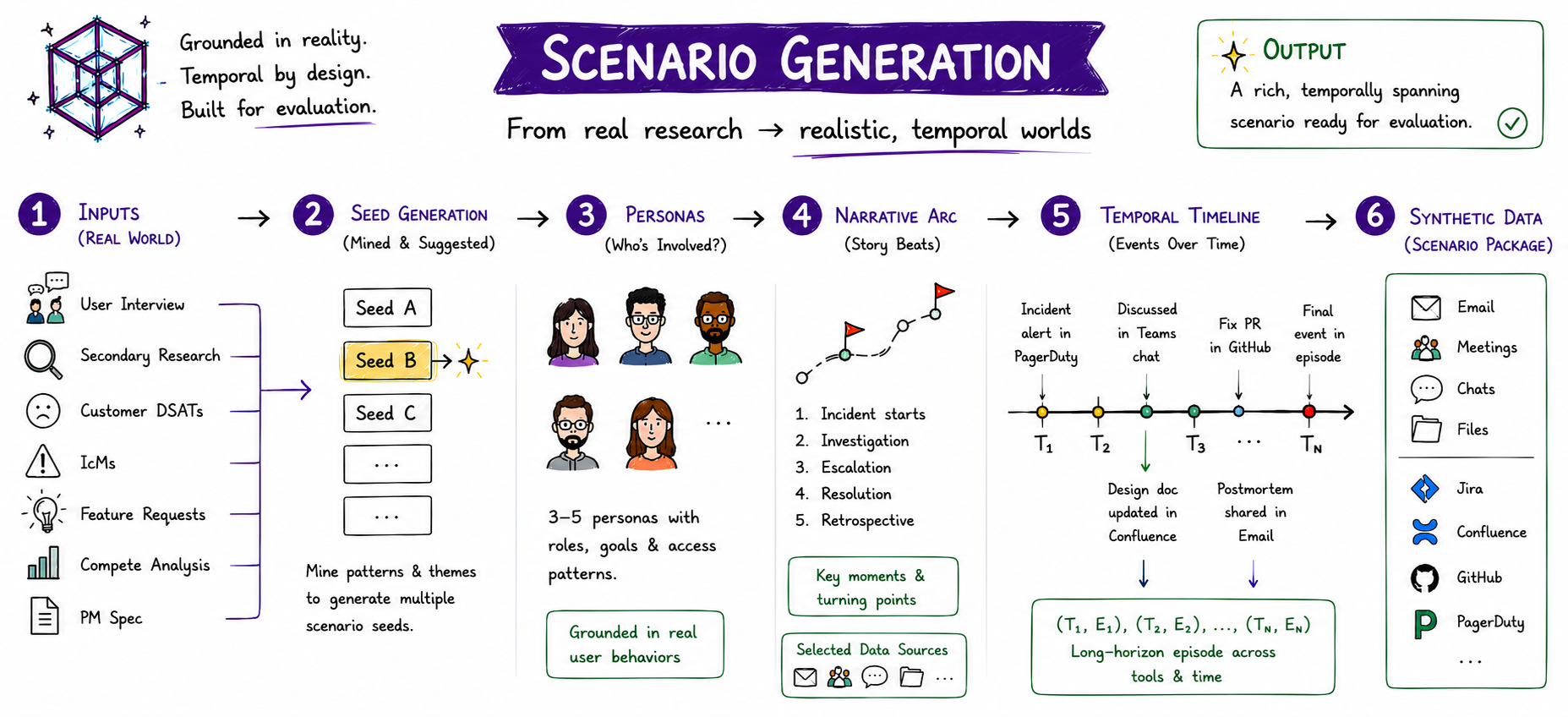}
\caption{Flow 1 (scenario generation): real-world research is mined into grounded, temporally-rich seeds, then expanded under approval gates into personas, a narrative arc, a temporal timeline of events across apps, and full-fidelity synthetic artifacts---a self-contained, replayable scenario package.}
\label{fig:gen}
\end{figure*}

An evaluation world is only as useful as it is believable; an agent that aces a contrived world tells you nothing. The expensive part is \emph{believability}---a coherent episode in which people, conversations, tickets, and code changes reference one another consistently \emph{over time}. The system makes that cheap two ways: it grounds the world in real material, and it generates it in an order that forces coherence (Fig.~\ref{fig:gen}).

\textbf{Research grounding and seed mining} (Fig.~\ref{fig:arch}). The system ingests a corpus of real artifacts a product team already has---interview notes, support tickets, incident write-ups, dissatisfaction reports---and mines \emph{recurring} situations across them into a handful of concrete scenario \emph{seeds}. Grounding in real material keeps the world representative of genuine pain rather than an author's imagination, and mining \emph{across} documents matters because the situations worth evaluating usually appear as a pattern spanning several sources, not any one.

\textbf{Gated scenario generation.} A seed is expanded one reviewed step at a time, and the order is a deliberate design choice, not convenience: \emph{personas} first (who is involved, and crucially \emph{what each is allowed to see}), then a \emph{narrative arc} (the cause-and-effect spine---what triggers what, and the turning points), then a \emph{timeline} placing each beat at a concrete timestamp across apps, and finally the \emph{artifacts} (emails, chat threads, tickets, pull requests, documents) those events produce. Jumping straight from people to a timeline, skipping the arc, reliably yields a bag of unrelated events no human would mistake for a real episode---the arc is the connective tissue. The result is the \emph{Ground-Truth World} of Fig.~\ref{fig:arch}---one omniscient record of the episode in which every artifact is stamped with the time it came into being and the set of personas permitted to see it, so \emph{time} and \emph{visibility} are first-class from the very first byte.

\textbf{Connector temporal augmentation.} Apps differ in how their records age. To onboard one---or a whole shipped \emph{package} of skills, plugins, and connectors---the system reads only its data \emph{schema} and infers a \emph{temporal description}: which fields carry time, which change, which become meaningful only after a prerequisite (a ``resolved-at'' time means nothing until resolution), which lists grow, and which are lifecycle states (a ``status''). It does this with cheap deterministic checks (spotting timestamp-typed fields and nested lists by their shape) plus a language model for the judgment calls. A new connector thus goes from ``just a schema'' to ``replayable through time'' in minutes, with no temporal-modeling expertise. As a concrete example, an incident record carries deterministic fields like a \texttt{timeline[]} that only grows and a \texttt{resolved\_at} that is empty until resolution, alongside a semantic \texttt{status} that must be \emph{derived} (triggered\,$\rightarrow$\,acknowledged\,$\rightarrow$\,resolved) and a free-text \texttt{description} that may have to be reworded (Appendix~A's incident schema illustrates this).

\section{Flow 2: Stepping Into the World at Any Instant}

\begin{figure*}[t]
\centering
\includegraphics[width=0.8\textwidth]{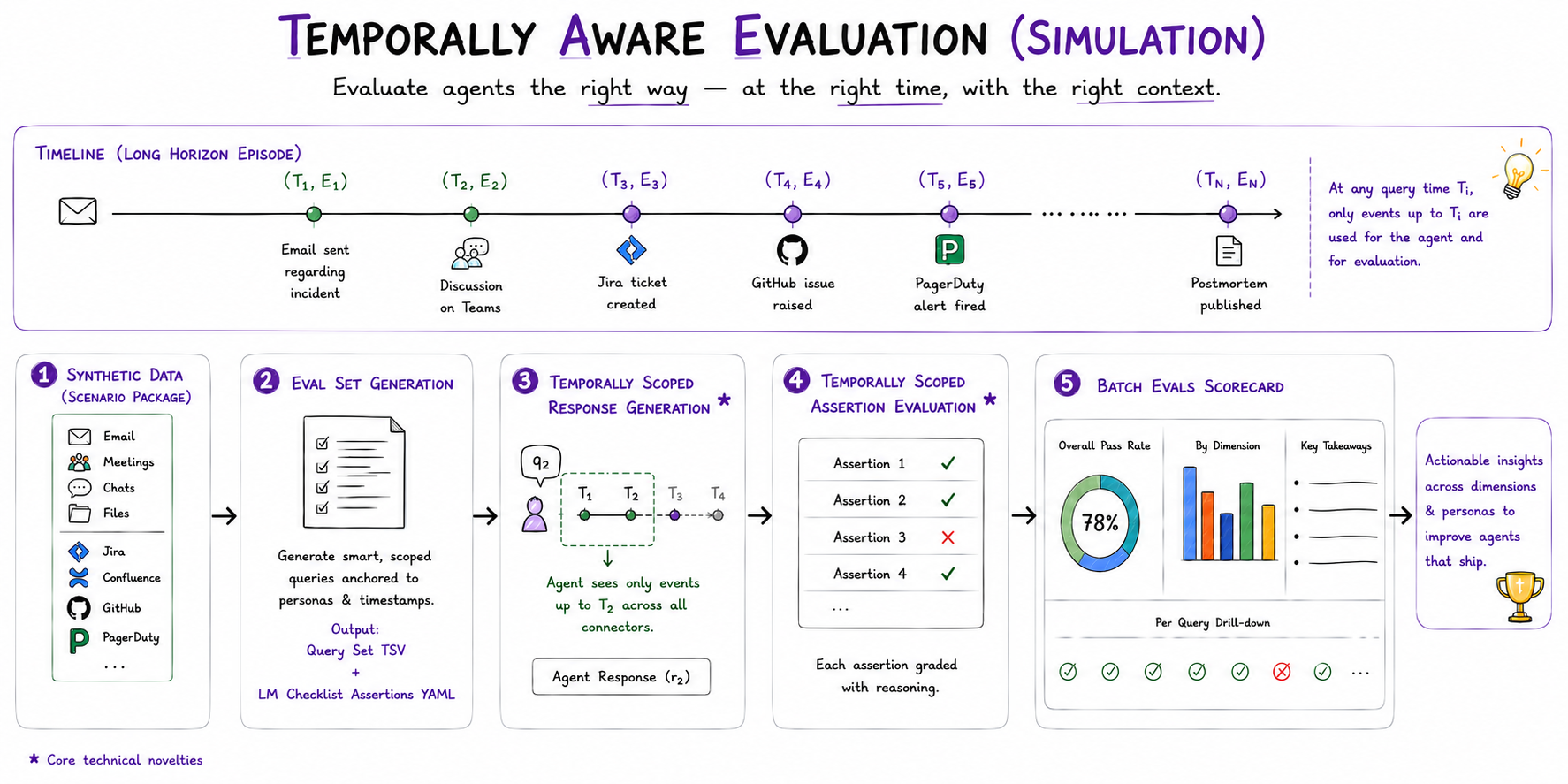}
\caption{Flow 2 (temporally-aware evaluation): from the replayable scenario package, evaluation questions and expectations are generated and anchored to a person and an instant; for each, the world is reconstructed at that instant by cache lookup and merge, the agent under evaluation is run through the adapter, and its answer is graded against the expectations to produce a scorecard.}
\label{fig:eval}
\end{figure*}

The hard problem of replay is simply stated: given the omniscient end-state and a target instant $t$, reconstruct each record \emph{exactly as it stood at $t$}---undoing everything that had not happened yet, including changes buried \emph{inside} the record---and then show a given person only what they were allowed to see.

\textbf{A two-stage rebuild.} Let the episode's events occur at instants $t_1 < t_2 < \dots < t_n$. A record $a$ has a birth time $\tau(a)$, a set $\mathrm{acc}(a)$ of people allowed to see it, and a final state $a^{*}$. Rebuilding $a$ as of $t$, written $\rho(a,t)$, runs in two stages,
\[
\rho(a, t) \;=\; \Lambda\big(D(a, t),\, t\big).
\]
The first stage $D$ is \emph{deterministic and cheap}: it drops sub-items dated after $t$ (later comments, later timeline entries), blanks any value whose own timestamp is in the future, blanks values that only make sense after a now-blanked prerequisite (the duration of a still-unresolved incident), and caps ``last-updated'' to $t$. The second stage $\Lambda$ is \emph{interpretive and used sparingly}: it runs only when $D$ actually changed the record, and only for the few fields whose past value cannot be read off the clock---a lifecycle \emph{status} (was the incident ``triggered'' or ``acknowledged'' at $t$?), or free text that mentions later events and must be reworded to what was known at $t$. This split is the key engineering decision: it keeps the bulk of the work exact, fast, and rule-based, and spends a language model only where judgment is genuinely unavoidable.

\textbf{Reconstruction precompute.} A language model in stage $\Lambda$ is both slow and \emph{nondeterministic}---fatal if it runs \emph{while grading}, because scores would wobble run to run. The decisive observation is that the only instants anyone will ever query are the finite $t_1,\dots,t_n$ on the timeline. So this stage computes every rebuild \emph{once}, ahead of time, and stores only its small \emph{difference} from the final state---the \emph{Precomputed Point-in-Time Differences} of Fig.~\ref{fig:arch},
\[
\delta(a, t_i) \;=\; \rho(a, t_i) \ominus a^{*},
\]
tiny because most of a record never changes. Identical differences are shared, and a fingerprint of the ground truth invalidates the cache if the world is ever edited.

\textbf{Replay is a lookup.} At grading time, the \emph{Point-in-Time Reconstruction Engine} (Fig.~\ref{fig:arch}) assembles the world person $P$ saw at $t$ by pure data operations---no model involved:
\[
\Sigma(t, P) \;=\; \{\, a^{*} \oplus \delta(a, t) \;:\; \tau(a) \le t,\; P \in \mathrm{acc}(a) \,\}
\]
---take every record that existed by $t$, merge its precomputed difference, and keep only those $P$ may see. Because nothing nondeterministic runs here, two evaluation runs over the same world are \emph{byte-identical}: reproducibility, the bedrock of regression evaluation, is guaranteed by construction rather than hoped for.

\textbf{Grading any agent.} Because the generator is omniscient, \emph{Query and Expectation Generation} writes the evaluation questions and their \emph{expected answers} directly---each tied to a person and instant---so grading needs no separate ``what was the right answer?'' lookup. The \emph{Agent-Agnostic Adapter} then hands the rebuilt world and the time- and visibility-scoped app tools~\cite{mcp} to whatever agent is under evaluation---a Copilot, a declarative agent, an Azure AI Foundry agent~\cite{foundryevals}, or a bespoke app---and records its reply; an \emph{Evaluator} grades that reply against the expected answers. Everything downstream is shared, so swapping the agent is a one-line change, and two competing agents can be judged on the \emph{identical} evolving world. Figure~\ref{fig:eval} summarizes this flow end-to-end.

\section{Design Principles and Early Lessons}

Four decisions did the heavy lifting; we offer them as transferable lessons for evaluating any nondeterministic, data-driven software.

\begin{itemize}
\item \textbf{Push nondeterminism out of the measured path.} Every model call that could move a score is run \emph{before} grading, into the precompute; what remains at evaluation time is pure data manipulation. Reproducibility then comes for free---far easier to engineer in than to retrofit.
\item \textbf{Generate in an order that forces coherence.} Coherence of synthetic data is not a post-hoc check but a consequence of generation order: persona\,$\rightarrow$\,arc\,$\rightarrow$\,timeline\,$\rightarrow$\,artifact bakes cause-and-effect in, where flat generation cannot.
\item \textbf{Separate ``cheap and exact'' from ``careful and rare.''} Most of point-in-time reconstruction is mechanical and should never touch a model; a small, clearly delimited remainder genuinely needs one. Drawing that line explicitly bought both speed and trust.
\item \textbf{Make the world agent-agnostic.} Decoupling the world from the agent through one adapter turns ``evaluate our agent'' into ``evaluate any agent on our world''---what makes fair comparison and cross-team reuse possible.
\end{itemize}

Early use on enterprise agents bore out the payoff: one authored episode yielded dozens of point-in-time, per-person evals; onboarding a new app fell from days to minutes; and batch runs were exactly reproducible. The main open difficulty is calibrating stage $\Lambda$---how aggressively it should reword free text per app, and how to \emph{validate} that a rebuilt past is faithful---which we turn to next.

\section{Discussion}

\textbf{Where this sits.} Table~\ref{tab:diff} contrasts the system with how agents are evaluated today. The capability that most sets it apart is evaluating always-on, event-driven agents~\cite{scout}: only a system that can step \emph{back into} an episode can place such an agent at the instant just after a change and ask whether it reacted---something a static snapshot can never do.

\begin{table}[h]
\small
\centering
\caption{Our system versus existing agent-evaluation approaches.}
\label{tab:diff}
\begin{tabular}{@{}p{3.5cm}ccc@{}}
\toprule
\textbf{Capability} & \textbf{Static} & \textbf{Sim.} & \textbf{the system} \\
 & \textbf{tenant} & \textbf{bench.}~\cite{taubench,workbench,theagentcompany} & \\
\midrule
Evaluate at any moment & --- & --- & \checkmark \\
Correct \emph{nested} record state & --- & --- & \checkmark \\
Reproducible model-derived state & n/a & n/a & \checkmark \\
Temporal behavior from a schema & --- & --- & \checkmark \\
Machine-generated expectations & manual & manual & \checkmark \\
Any agent via one adapter & --- & --- & \checkmark \\
\bottomrule
\end{tabular}
\end{table}

\textbf{Reliability and reproducibility.} Moving all model-based interpretation into the precompute keeps the grading path deterministic; the closest academic system for continuous enterprise benchmark generation~\cite{contbench} explicitly cannot handle temporally-evolving connector snapshots, which is exactly what the difference cache provides. Temporal knowledge graphs and time-travel retrieval~\cite{zep,timetravel} reconstruct history for agent \emph{memory} or document retrieval, not to generate an access-scoped world for grading. The open calibration question is per-app: how aggressively $\Lambda$ should rewrite free text, and how to validate that a rebuilt world is faithful to the moment.

\textbf{Generalization and what is next.} The persona/timeline/artifact model and the schema-to-temporal-description mechanism assume nothing Microsoft-specific; any platform with evolving, multi-app, access-controlled records is a candidate. Two forward directions stand out: \emph{what-if} (counterfactual) evaluation, forking the world at a decision point to compare alternatives, and a \emph{self-learning flywheel}, using observed failures to generate new people, moments, and questions that target weak spots---a step toward high-quality self-learning for enterprise agents.

\section{Conclusion}

The system reframes offline agent evaluation from grading against a frozen end-state to replaying an access-scoped world at any moment in time. By pairing research-grounded, persona-driven \emph{generation} with deterministic, precomputed \emph{point-in-time replay}, it turns one synthetic episode into a faithful, reproducible, time-aware evaluation suite for the cross-app agents, skills, and connectors now shipping into the enterprise.

\bibliographystyle{ACM-Reference-Format}

\section*{Appendix A: Worked Example}

A payments-checkout service degrades, and the episode unfolds over two hours across four apps: an \emph{incident tool} (the alert and its growing timeline), a \emph{Teams} channel (responders coordinate), \emph{email} (a stakeholder update), and a \emph{code repository} (the fix, as a pull request). Three people are involved, with different visibility: the on-call engineer \emph{Rohan} and his manager \emph{Vikram} can see the incident, the Teams channel, and the repository; the product manager \emph{Priya} is kept informed only through the email thread. Generation timestamps every event: $t_1{=}18{:}42$ the incident is triggered; $t_2{=}18{:}47$ Rohan acknowledges it and posts in Teams; $t_3{=}19{:}05$ he adds a triage note correlating a recent configuration change; $t_4{=}20{:}10$ a fix is opened as a pull request; $t_5{=}20{:}38$ it is merged; $t_6{=}20{:}55$ the incident is resolved with a root cause and a postmortem link; a summary email reaches Priya at 21:00.

We evaluate at $t_3 = 19{:}05$. Rohan asks the agent: \emph{``What is the current status of the incident I'm leading, what is on its timeline so far, and is a fix already in progress?''} A correct answer must reflect the world \emph{as it stood at 19:05}---across the incident tool, Teams, and the repository---and only what Rohan may see. Table~\ref{tab:schema} gives the incident connector's temporal schema and the resulting rebuild of the incident record at 19:05: stage $D$ resolves the deterministic fields, stage $\Lambda$ derives the semantic ones.

\begin{table*}[t]
\small
\centering
\caption{The incident connector's temporal schema---deterministic vs.\ semantic fields and the rule for each---and the rebuild of one incident record at 19:05. Deterministic fields are resolved by stage $D$; semantic fields are derived by stage $\Lambda$.}
\label{tab:schema}
\begin{tabular}{@{}p{1.9cm}p{1.4cm}p{5.4cm}p{3.1cm}p{3.4cm}@{}}
\toprule
\textbf{Field} & \textbf{Kind} & \textbf{Temporal rule} & \textbf{Final state $a^{*}$} & \textbf{Rebuilt at 19:05} \\
\midrule
\texttt{timestamp} & determ. & record exists only if \texttt{timestamp}~$\le$~as-of & 18:42 & 18:42 (kept) \\
\texttt{timeline[]} & determ. & drop child entries whose \texttt{time}~$>$~as-of & 10 entries & 3 entries \\
\texttt{resolved\_at} & determ. & null if its value~$>$~as-of & 20:55 & null \\
\texttt{root\_cause} & determ. & null when \texttt{resolved\_at} is null (\emph{depends-on}) & full RCA text & null \\
\texttt{status} & semantic & derive the lifecycle from the trimmed timeline: triggered~$\rightarrow$~acknowledged~$\rightarrow$~resolved & resolved & acknowledged \\
\texttt{description} & semantic & reword if it cites events after as-of & ``\ldots merged fix; service restored 20:55'' & ``error spike under investigation; config change suspected'' \\
\bottomrule
\end{tabular}
\end{table*}

Two effects are worth highlighting. \textbf{Cross-app:} at 19:05 the fix does not yet exist---the pull request opens at 20:10---so a faithful world contains no merged code, and the rebuilt \texttt{description} must drop its reference to that not-yet-opened fix, a value that originates in a \emph{different} app. \textbf{Visibility:} $\Sigma(19{:}05, \text{Rohan})$ contains this incident (with its three-entry timeline) and the Teams messages up to 19:05, but Priya---who sees only the email thread, whose first message arrives at 21:00---would see nothing of the incident at this instant. The expectation the generator wrote for the question---\emph{``status is acknowledged; the timeline shows the configuration-change triage note; no root cause; no fix in progress yet''}---then grades the answer: a reply that says ``resolved'' or mentions the merged fix fails, because it is quoting a future that had not happened.

\section*{Appendix B: Terminology}

The paper uses several AI-extensibility terms with distinct, easily-confused meanings; Table~\ref{tab:glossary} fixes how each is used. They are listed in the order each builds on the previous.

\begin{table*}[t]
\small
\centering
\caption{Terminology: the AI-extensibility ecosystem this work plugs into.}
\label{tab:glossary}
\begin{tabular}{@{}p{2.4cm}p{13.6cm}@{}}
\toprule
\textbf{Term} & \textbf{Meaning, with an example} \\
\midrule
App / enterprise app & An application where enterprise work happens and data lives. \emph{E.g.,} Outlook, Teams, Jira, GitHub, ServiceNow. \\
Tool & A single callable function, with a defined input and output, that a model can invoke. \emph{E.g.,} a \texttt{search\_issues} function. \\
Agent & An autonomous AI system, configured with scoped instructions, that plans and calls \emph{tools} to accomplish a task. \emph{E.g.,} Microsoft 365 Copilot, an Azure AI Foundry agent. \\
MCP & Model Context Protocol --- an open standard for exposing an \emph{app}'s \emph{tools} and data to any compliant \emph{agent}. \\
Connector & Brings an external \emph{app}'s data and actions to an \emph{agent}. \emph{Sync} (graph): ingests and indexes the app's content for local search (read only). \emph{Federated}: reached live over \emph{MCP}; reads and writes. \\
Skill & A packaged set of instructions and resources that guides an \emph{agent} through a domain or multi-step task. \emph{E.g.,} a discounted-cash-flow finance skill. \\
Plugin & A distributable package that bundles \emph{skills}, \emph{tools}, and \emph{connectors} into one installable unit extending an AI \emph{app}. \emph{E.g.,} an analytics plugin. \\
Package & A shipped bundle of \emph{skills}, \emph{plugins}, and the \emph{connectors} they need --- released and evaluated as a unit. \\
\bottomrule
\end{tabular}
\end{table*}

\end{document}